\documentclass[aps,pra,twocolumn,showpacs,showkeys,groupedaddress]{revtex4}
\usepackage{graphicx}
\usepackage{graphics}
\usepackage{color}  
\usepackage{ulem} 
\usepackage{upgreek}

\newcommand{\etal}{{\it et al.}~}

\newcommand{\fprt}{f({\mathbf{p}}, {\mathbf{r}},t)}

\begin{document}
\title{Vortex reconnections in atomic condensates at finite temperature}
\author{A.~J. Allen$^1$}\email{Joy.Allen@ncl.ac.uk}
\author{S.~Zuccher$^2$}
\author{M.~Caliari$^2$}
\author{N.~P. Proukakis$^1$}
\author{N.~G. Parker$^1$}
\author{C.~F. Barenghi$^1$}
\affiliation{$^1$Joint Quantum Centre (JQC) Durham-Newcastle, School of Mathematics and Statistics, 
Newcastle University, Newcastle upon Tyne NE1 7RU, England, UK.\\
$^2$Dipartimento di Informatica, Universit{\`{a}} di Verona, Ca' Vignal 2, Strada Le Grazie 15,
37134 Verona, Italy}%

\date{\today}
\begin{abstract}
The study of vortex reconnections is an essential ingredient of understanding superfluid turbulence, a phenomenon recently also reported in trapped atomic Bose--Einstein condensates.  In this work we show that, despite the established dependence of vortex motion on temperature in such systems, vortex reconnections are actually temperature independent on the typical length/time scales of atomic condensates.  Our work is based on a dissipative Gross-Pitaevskii equation for the condensate, coupled to a semiclassical Boltzmann equation for the thermal cloud (the Zaremba--Nikuni--Griffin formalism).
Comparison to vortex reconnections in homogeneous condensates further show reconnections to be insensitive to the inhomogeneity in the background density.

\end{abstract}

\maketitle


In classical hydrodynamics, reconnections of stream lines, vortex lines and
magnetic flux tubes change the topology of the flow and contribute to 
energy dissipation. In quantum fluids, vorticity is not a continuous field,
but is concentrated in discrete vortex lines of
quantised circulation; their reconnections are therefore isolated, dramatic
events.  Individual quantum reconnections have
been recently visualized \cite{Paoletti2010} in superfluid $^4$He. 
The role of vortex reconnections in the dynamics
of quantum turbulence \cite{Barenghi-Skrbek-Sreeni2014}
in superfluid $^4$He, superfluid $^3$He and
atomic Bose--Einstein condensates (BECs) is currently debated. For example, one would like
to understand their contribution to acoustic dissipation of kinetic
energy \cite{Leadbeater2001}, 
their role in the proposed bottleneck \cite{Lvov-Nazarenko-Rudenko2008}
between a semi--classical
Kolmogorov cascade at small wavenumbers and a quantum Kelvin wave cascade
at large wavenumbers, and the possibility of a cascade of vortex
rings scenario \cite{Kursa,Kerr}. The increasing ability to 
imprint~\cite{leanhardt_gorlitz_02}, generate~\citep{raman_aboshaeer_01,anderson_haljan_01,weiler_neely_08,freilich_bianchi_10}, manipulate~\cite{davis_carretero_09}
and directly image~\citep{madison_chevy_00,freilich_bianchi_10} vortices makes atomic condensates ideal systems to study vortex reconnection events~\citep{baggaley_14}.  This problem is of particular interest in light of recent experiments regarding quantum turbulence and vortex dynamics in both two~\cite{neely_bradley_13,wilson_samson_13,kwon_moon_14} and three dimensions~\cite{henn_seman_09} (for a review on progress in two and three dimensions see, e.g.~\citep{white_anderson_14}).

Since many experiments are performed at relatively high temperatures,
 i.e. large fractions of the critical temperature, $T_c$, a natural question to ask is 
whether thermal excitations affect vortex reconnections.  A recent experiment visualising vortex reconnections in superfluid $^4$He, suggests that this is not the case~\citep{Paoletti2010}.  However, previous
investigations have shown that the presence of a thermal cloud significantly changes the motion of vortices in harmonically trapped condensates~\cite{fedichev_shlyapnikov_99,
schmidt_goral_03,duine_leurs_04,jackson_proukakis_09,
rooney_bradley_10,allen_zaremba_13}.

In this paper
we present results of an investigation of vortex reconnections
in finite--temperature trapped Bose--Einstein condensates. We model the problem
in the context of the Zaremba--Nikuni--Griffin (ZNG) formalism \cite{zaremba_nikuni_99,griffin_nikuni_book_09}, where the Gross-Pitaevskii
equation (GPE) is generalized by the inclusion of the thermal
cloud mean field, and a dissipative or source term which is
associated with a collision term in a semiclassical Boltzmann
equation for the thermal cloud. 
The main feature of this model is that the condensate and thermal cloud
interact with each other self--consistently; for 
a strongly nonlinear dynamical event like a vortex reconnection,
a simpler and less accurate approach may give misleading answers.

The governing ZNG equations are  

\begin{equation}
i \hbar \frac {\partial \phi}{\partial t} = 
\left( -\frac{\hbar ^2 }{2m} \nabla^2
+ V_{\mathrm{ext}} + g \left[n_c  
+ 2\tilde n \right]- iR \right)\phi \;, 
~\label{eq:zng_gpe}
\end{equation}

\noindent
and

\begin{equation}
\frac{\partial f}{\partial t} + \frac{\bf{p}}{m} \cdot \nabla_{\bf{r}} f
- (\nabla_{\bf{r}}U_{\rm{eff}}) \cdot
(\nabla_{\bf{p}}f)= C_{12}+ C_{22}.\\
\label{eq:zng_qbe}
\end{equation}

\noindent
In this formalism $\phi=\phi(\mathbf{r},t)$ is the condensate wavefunction,
$f=f(\mathbf{r},\mathbf{p},t)$ is the phase-space distribution function
of thermal atoms, $V_{\rm ext}=m \omega^2 r^2/2$ is the harmonic
potential which confines the atoms (assumed, for simplicity, to be spherically-symmetric), $\omega$ is the trapping frequency, $m$ the atomic mass, and $g=4 \pi \hbar^2 a_{\rm s}/m$, with $a_{\rm s}$ being the {\it s}-wave scattering length.  Equation~(\ref{eq:zng_gpe}) generalises the GPE for a $T=0$ condensate by the addition of the thermal cloud
mean--field potential $2 g \tilde n$ 
and the dissipation/source term $-iR(\mathbf{r},t)$. The condensate
density is $n_c(\mathbf{r},t)=\vert \phi(\mathbf{r},t) \vert^2$ and
the thermal cloud density is recovered from $f(\mathbf{r},\mathbf{p},t)$ 
via an integration over all momenta, 
$\tilde n ({\bf{r}}, t) = (2\pi
\hbar)^{-3} \int d {\bf{p}}f({\bf{p}},{\bf{r}},t)$.  
The mean-field potential acting on the thermal cloud is 
$U_{\rm{eff}} = V_{\rm {ext}}(\mathbf{r}) +2g[n_c({\mathbf{r}},t) +
\tilde n({\mathbf{r}},t)]$.
The quantities $C_{22}[f]$ and $C_{12}[\phi,f]$ 
are collision integrals defined in
Refs.~\citep{zaremba_nikuni_99,griffin_nikuni_book_09} (which contain further details of the model).
$C_{22}$ describes the 
redistribution of thermal atoms as a result of two--atom collisions
(as in the usual Boltzmann equation), while 
$C_{12}$, which is related to $-iR$ via $
R({\bf{r}}, t) = {\hbar}/{(2n_c({\bf{r}}, t))} \int {d {\bf{p}}}/{(2\pi \hbar )^3}
C _ {12}[\phi({\bf{r}}, t),f({\bf{p}},{\bf{r}},t)] $, describes the change in the phase-space
distribution function $\fprt$ resulting from particle-exchange collisions between thermal atoms
and condensate atoms.

Before solving the ZNG equations (\ref{eq:zng_gpe}) and (\ref{eq:zng_qbe}) 
numerically, we write them in dimensionless form,
using the harmonic oscillator's 
characteristic length $\ell = \sqrt{\hbar /m \omega}$ as the unit of distance, 
the inverse trapping frequency $\omega^{-1}$ as the unit of time, 
and $\hbar \omega$ as the unit of energy.
We choose experimentally realistic parameters:
$\omega = 2 \pi \times 150$Hz,  
${\tilde \mu}=\mu/(\hbar \omega) \approx 18$
where $\mu$ is the chemical potential,
and ${\tilde g}=g/(\ell \hbar \omega) \approx 6230$.  This corresponds to $\ell = 0.88 \upmu$m, $\omega^{-1} = 1.06$ms and $N_c \approx 75000$ $^{87}$Rb atoms. Throughout this work, in order to facilitate more direct comparisons, we keep this value of $N_c$ approximately constant even at temperatures above zero so that the effect of increasing temperature is to increase the number of thermal atoms in the system rather than depleting the condensate.   As well as solving the ZNG equations, we use another model to describe finite temperature effects.  A simple phenomenological extension of the GPE known as the dissipative Gross--Pitaevskii equation (DGPE, (Eq.~\ref{eq:dgpe})), is used for temperatures below $0.4T_c$, which is thought to be reasonable limit for its validity~\citep{choi_morgan_98}, while in the trapped case, for temperatures higher than this we use the ZNG equations.

First we consider what happens in the limit of zero temperature,
for which Eq.~(\ref{eq:zng_gpe}) with $iR = 0$ and $\tilde n  = 0$ reduces to the GPE, a model known to provide an accurate description of condensate dynamics for $T \ll T_c$, 
including collective modes and vortex dynamics  
~\citep{pitaevskii_stringari_book_03,pethick_smith_book_02}.  For the quantities considered here, we have confirmed that the results of the ZNG equations in the absence of any significant thermal cloud, agree with the results given by the GPE.  
The initial state of our simulation is the condensate containing a pair of straight line anti-parallel vortices aligned in the $z$--direction at locations $(x_0/\ell,y_0/\ell) = (-1,\pm1)$.  This state is formed by imaginary time propagation of the GPE while enforcing a $2 \pi$ winding of the phase of $\phi$ around the location of the cores.    

\begin{figure}
\centering{
\includegraphics[scale = 0.62]{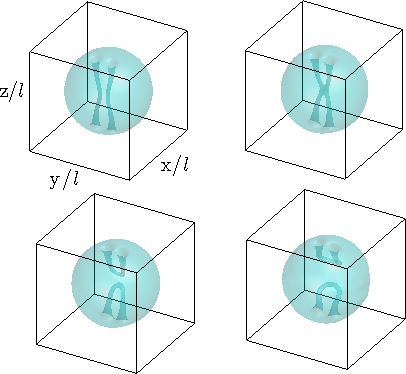}
}
\caption{Vortex reconnection in the $T=0$ trapped condensate.
Density isosurfaces of the dynamics of the initial anti--parallel vortex pair,
at times $t\omega = 0$ (top left), $t\omega = 1.19$ (top right), $t\omega =1.80$ (bottom left) 
and $t\omega =2.5$ (bottom right), according to the GPE.  
The isosurfaces are plotted at $40\%$ of the peak density. The range of the axes is $-6 \ell \le x, y, z \le 6 \ell$.}
\label{fig1}
\end{figure}

In the GPE model, the vortex core size is of the order of the condensate healing length, $\xi = \hbar/\sqrt{2m\mu}$.  For our assumed parameters, $\xi/\ell \approx 0.167$.  This is much smaller than the condensate radius, given approximately 
by the Thomas-Fermi radius, $R_{\rm{TF}}/\ell = \sqrt{2\tilde \mu}\approx6$.
To ensure that the
vortices reconnect in the central region of the condensate and away from its boundary, we perturb
their $y$ position along $z$ according to
$A[\cos(2\pi z/\lambda)]^6$ 
where $A/\ell = 0.5$ and $\lambda/\ell=20$.  
This initial condition is shown in the top left panel 
of Fig.~\ref{fig1}, where we have plotted a series of 
snapshots of the reconnection for the case of $T = 0$.  
It is apparent that the two vortices initially move as a pair in the $xy$ 
plane, traveling in the $x$ direction. The slight initial curvature
enhances the Crow instability~\citep{crow}: the vortices approach each other and
reconnect, creating two U--shaped cusps which lift and move away
from each other above and below the $xy$ plane (bottom right panel).

\begin{figure}[h!]
\begin{flushleft}
$~~~~~~~~~~~~~~T = 0~~~~~~~~~~~~~~~~~~~~~~~~~~~~T>0$
\end{flushleft}
\centering{
\includegraphics[scale = 0.75]{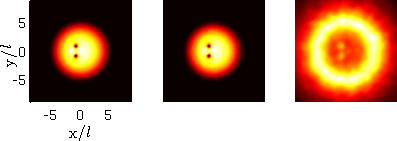}
\includegraphics[scale = 0.75]{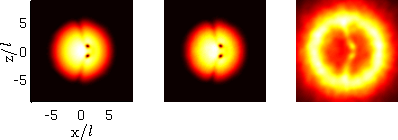}
\includegraphics[scale = 0.75]{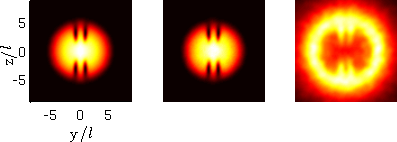}
}
\caption{Vortex reconnection in the trapped condensate at $T=0$ (left column) and $T/T_c=0.62$ (middle and
  right columns).
2D density plots from the ZNG model of the condensate (left and middle) and thermal
cloud (right).  Top: Before the reconnection,  $t\omega=0.3$,  on the $z/\ell=-2.5$ plane (rather
than the $z/\ell=0$ plane, in order to better distinguish the two vortex cores).
Middle and bottom:  After the reconnection, $t\omega = 2$, slices through the $xz$ plane for $y/\ell=0$
(middle row) and the $yz$ plane for $x/\ell=0$ (bottom row).
In the density scale, white corresponds to peak density and black to zero density.
Note the vanishing
condensate density in the vortex cores (left and middle column) and the concentration of
thermal atoms at the edge of the condensate and inside the cores (right column).}
\label{fig2}
\end{figure}

Simulations at various temperatures show that
the vortex reconnection proceeds essentially unaffected by the
presence of the thermal cloud, consistent with the findings of Ref.~\citep{Paoletti2010}.  To illustrate this we focus on the relatively high temperature case of $T/T_c=0.62$.  Figure~\ref{fig2} compares density slices of the $T=0$ condensate (left panel) and the $T/T_c=0.62$ condensate and thermal cloud
(middle and right panels), both before reconnection (top row) and after reconnection (middle and bottom rows).  It is apparent that thermal atoms are concentrated at the edge of the condensate and inside the vortex cores~\citep{allen_zaremba_13}, an effect of the mean--field repulsion from the condensate.  Importantly, over the time scale for the reconnection, we observe no difference in the vortex dynamics between the $T=0$ and the $T>0$ cases.

The above numerical simulations refer to the 
typical experimental situation
 where the condensate is larger, but not much larger, than
the vortex core size ($R_{\rm{TF}} \approx 36 \xi$).  In this case, and as evident in
Figs.~\ref{fig1} and \ref{fig2}, the reconnection region
is not far from the condensate outer surface.   This surface region can undergo oscillations, particularly in a turbulent condensate
\cite{white2010}, and interact with the vortices.  Moreover the surface region is where thermal atoms accumulate, and is likely to introduce relatively large frictional effects upon the vortices.  In order to more clearly extract out the role of finite temperature on reconnections it is therefore instructive to consider reconnections in a homogeneous (boundary-free) condensate.  

In the interest of computational ease and efficiency, we can use a simpler extension to the GPE which simulates thermal effects by the inclusion of a phenomenological damping parameter, called the dissipative Gross--Pitaevskii equation (DGPE)~\citep{choi_morgan_98,jackson_proukakis_08}, 
\begin{equation}
(i -\gamma) \hbar \frac {\partial \phi}{\partial t} = 
\left( -\frac{\hbar ^2 }{2m} \nabla^2
+ g \vert \phi \vert^2 - \mu \right) \phi.
~\label{eq:dgpe}
\end{equation}
The phenomenological damping parameter $\gamma$, which we assume
to be constant, has been used in a variety of systems to simulate
thermal effects (see e.g.~\citep{choi_morgan_98,proukakis_parker_04,tsubota_kasamatsu_02,madarassy_barenghi_08}).  For $\gamma=0$ this model reduces to the
GPE.  We solve Eq.~(\ref{eq:dgpe}) in a periodic box in the absence of an external potential.  It must be stressed that,
unlike the ZNG model, the DGPE
does not include the dynamics of the thermal cloud.

\noindent

Before solving Eq.~(\ref{eq:dgpe}) we write it in dimensionless
form using natural units based on the healing length 
$\xi=\hbar/\sqrt{2m \mu}$ and the time unit $\xi/c$.
As done by Zuccher \etal~\cite{zuccher_caliari_12}, we use a Fourier-splitting 
scheme where the Laplacian part is trivially solved in spectral space,
whereas the remaining part is exactly solved in physical space as suggested 
by Bao \etal~\cite{bao_etal_2006}.
We place
a pair of anti--parallel vortex lines in a computational 
box of size $-30 \xi\le x,y,z \le 30\xi$ at
position $(x_0/\xi, y_0/\xi) = (10, \pm 3)$. Again, to accelerate the reconnection and ensure it occurs away
from boundaries, the vortex lines are initially perturbed according to $A[\cos(2\pi z/\lambda)]^6$, now
with $A/\xi = 1$ and $\lambda/\xi = 120$  (chosen such that the  
the vortices are unperturbed at $z_{\rm{max}}$ and $z_{\rm{min}}$).  
The box size is choosen such that the vortex
length is comparable to the vortex length in the ZNG simulation.

\begin{figure}
\centering{
\includegraphics[scale = 0.16]{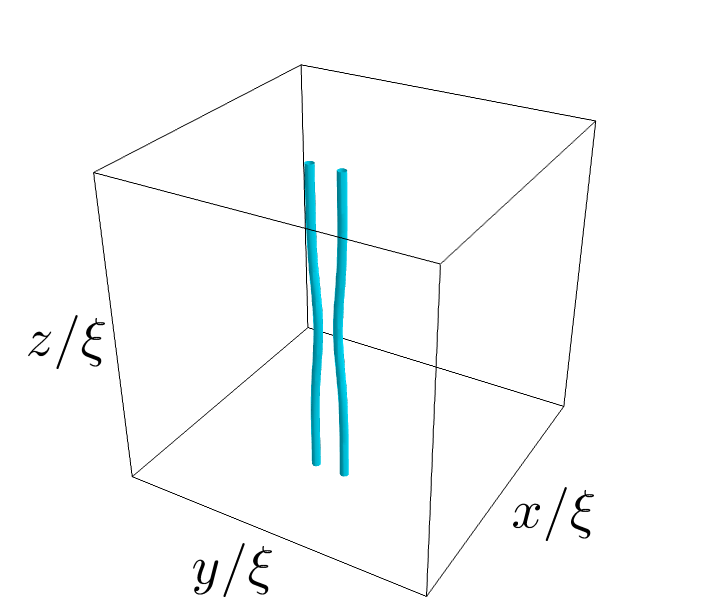}
\includegraphics[scale = 0.16]{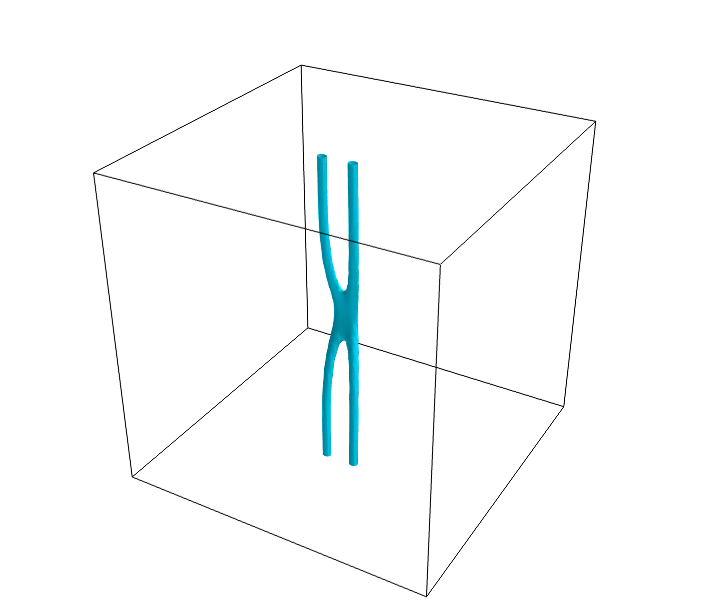}
\includegraphics[scale = 0.16]{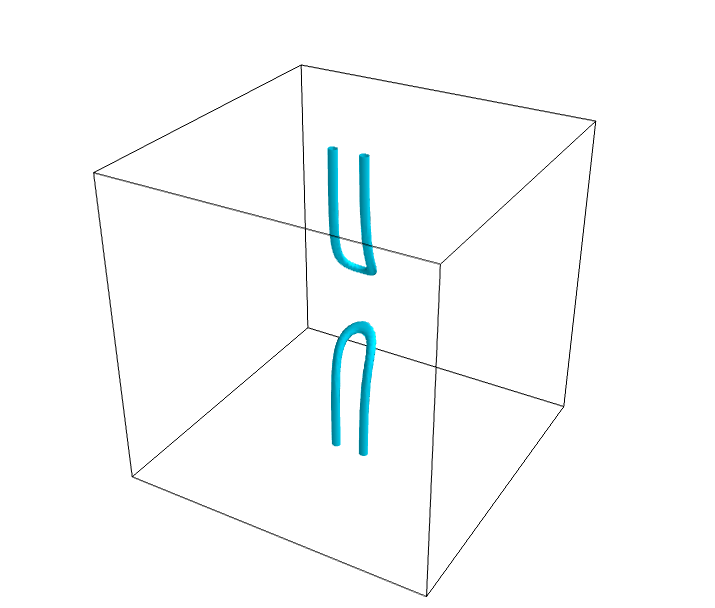}
\includegraphics[scale = 0.16]{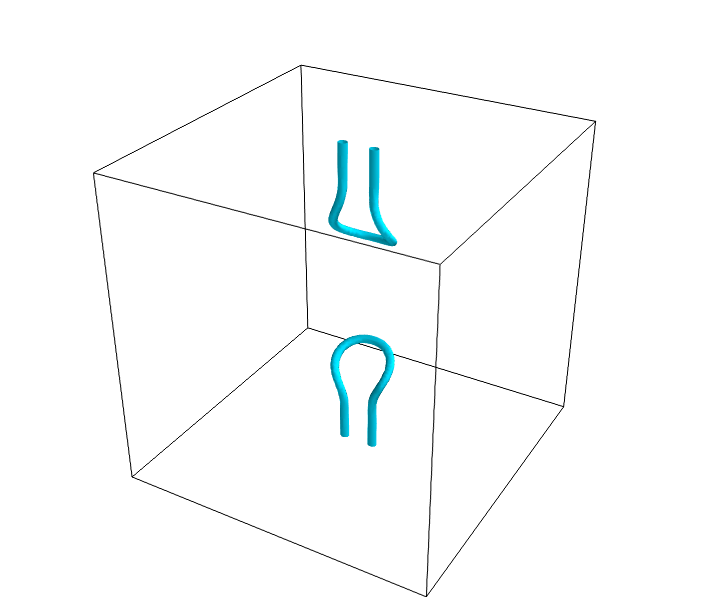}
}
\caption{Vortex reconnection in the homogeneous condensate at $T=0$.
Density isosurfaces showing the reconnection of the vortex-antivortex pair, according to the GPE,
at times $t/(\xi/c) = 0$ (top left), $t/(\xi/c) = 20$ (top right), $t/(\xi/c) =40$ (bottom left) 
and $t/(\xi/c) =60$ (bottom right).  
The isosurfaces are plotted at $20\%$ of the peak density. The range of the axes is $-30 \xi \le x, y,
z \le 30 \xi$.}
\label{fig:hom_iso}
\end{figure}

The initial configuration and subsequent evolution of the vortex pair is depicted in Fig.~\ref{fig:hom_iso} for $\gamma = 0$, corresponding to $T=0$.  The vortex reconnection proceeds in much the same way as for the trapped condensate shown in Fig.~\ref{fig1}.  We repeat the simulation for $\gamma = 0.03$ (corresponding to a temperature of approximately $0.4T_c$ ~\citep{choi_morgan_98,penckwitt_ballagh_02,kasamatsu_tsubota_03}) and again note that the reconnection proceeds essentially unchanged despite the presence of dissipation in the system.  

To monitor the vortex reconnections more precisely than ``by eye", we consider the minimum distance between the vortex lines, 
$\delta(t)$.  We extract the position of the vortex core by finding the grid points where the density is a local minimum
and about which the phase changes by $2\pi$~\citep{allen_parker_14}. The time--dependence of this quantity (before and after
the reconnection) was experimentally observed in superfluid $^4$He, and predicted theoretically for superfluids based on the GPE ($T=0$) \cite{zuccher_caliari_12} and for ordinary viscous fluids based on the classical Navier--Stokes equation~\citep{hussain_duraisamy_11}.  
To enable comparison of $\delta(t)$ between the homogeneous and trapped systems, we must convert between harmonic trap units (based on the harmonic
oscillator length and frequency) and natural units (based
on the healing length and the chemical potential).  The conversion for length from harmonic oscillator units to natural units is given by $x' =
{\tilde{x}} \ell/\xi$ and for time is $t' =  c/(\omega \xi) \tilde{t}$ where we used a tilde to denote the quantity in harmonic trap units and a prime for natural units (see footnote~\footnote{For harmonic oscillator units, length and time are defined as $\tilde x = x/\ell$ and $\tilde t =
\omega t$ respectively (where the tilde represents the quantity in harmonic oscillator
dimensionless units).  Similarly for natural units we define length and time as $x' = x/\xi$ and $t'
=  t/(\xi/c)$ respectively, where $c = \sqrt{\mu/m}$ 
is the sound speed (the prime denotes the quantity in natural dimensionless units)} for more details).  For the remainder of 
this article we express all quantities in natural units.  

\begin{figure}
\centering{
\includegraphics[scale = 0.4]{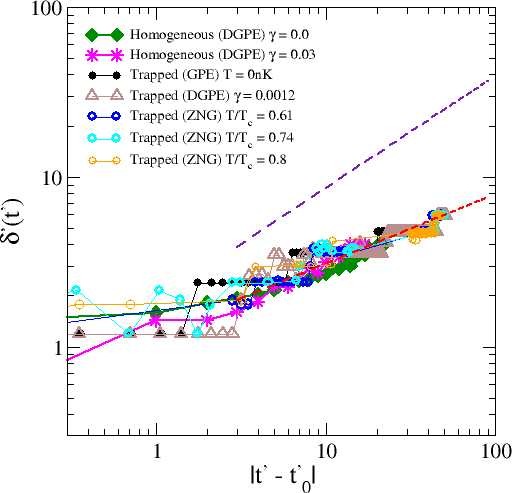}
\includegraphics[scale = 0.4]{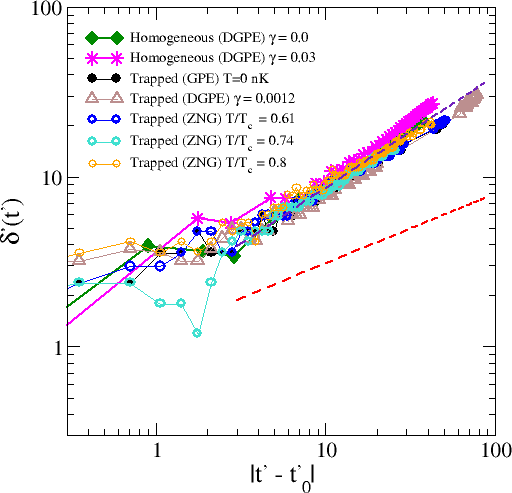}}
\caption{Minimum distance between vortex line, $\delta'(t')$,
before (top) and after (bottom) the vortex-antivortex reconnection,
computed for a harmonically trapped condensate and a homogeneous 
condensate at different values of $T/T_c$ and $\gamma$ (both the homogeneous and trapped values of $\gamma$ correspond to $T\approx 0.4T_c$),
including the limiting cases $T=0$ and $\gamma=0$. 
The dashed lines are fits according to Eq.~(\ref{eq:fit}) with parameters $\nu = 0.41\pm0.02$ (line of best fit before reconnection, red, dot-dashed) and $\nu = 0.66\pm0.02$ (line of best fit after reconnection, indigo, dashed) with corresponding $\kappa$ values, for the data considered here, of $1.36$ and $1.88$ respectively.
}
\label{fig:fig5}
\end{figure}


Figure~\ref{fig:fig5} compares $\delta'(t')$
computed using the ZNG model (trapped condensate) and the
DGPE (homogeneous condensate) before (top) and after the reconnection 
(bottom).  For completeness, we have also carried out DGPE simulations, within the presence of a trap for the value $\gamma = 0.0012$ which corresponds to a temperature of approximately $0.4T_c$ (as used for the homogeneous case).  We find that $\delta'(t')$ scales as

\begin{equation}
\delta' (t') = \kappa|t' - t_0'|^{\nu}
\label{eq:fit}
\end{equation}

\noindent
where $t_0$ is the time at which the reconnection takes place (defined as when $\delta'(t') = 0$),
and  $\kappa$ and $\nu$ are fitting parameters. It is apparent that the results
are essentially independent of the model, the temperature and the presence/absence of trapping. The best fit to the
parameter $\nu$ before the reconnection is $\nu = 0.41\pm0.02$ and after the reconnection is $\nu = 0.66\pm0.02$.  

Our results compare well with the exponents
$\nu=0.39$ ($t<t_0$) and $\nu=0.68$ ($t>t_0$)
reported by Zuccher \etal~\cite{zuccher_caliari_12}
over a wide range of initial angles between
the vortex lines, computed for $T=0$ (GPE) in a homogeneous condensate. 
It is interesting to remark that
viscous reconnections of the Navier--Stokes equation display a similar time
asymmetry \cite{hussain_duraisamy_11} (the largest $\nu$ being that
after the reconnection, as in a quantum fluid).
Zuccher \etal~\cite{zuccher_caliari_12}
argued that the time asymmetry is due to acoustic
emission: part of the kinetic energy of the vortices is transformed into
sound waves which radiate to infinity, in analogy with viscous 
dissipation in an ordinary fluid which turns kinetic energy into heat.
Indeed, if one uses the Biot--Savart law (an incompressible
model) to monitor the behaviour of vortices just before and after the
reconnections, one finds $\nu=0.5$ for both $t<t_0$ and $t>t_0$.
Paoletti \etal~\citep{Paoletti2010} observed individual quantum reconnections in
superfluid $^4$He experiments, reporting the exponent $\nu=0.5$ averaged over
all $t<t_0$ and $t>t_0$ data. Above all, Paoletti \etal
 did not notice any temperature
dependence, which is consistent with our findings.  

We stress that, a priori, one would not apply the Biot--Savart
model to a vortex in a small atomic condensate, as the vortex core is 
not negligible, particularly in a reconnection, when
two vortex cores collide.  However, we may gain some insight to the temperature--independence of the reconnecting behaviour from it.
In the Biot--Savart model~\cite{Hanninen}
the vortex is  a three-dimensional space curve 
${\bf{s}}\equiv{\bf{s}}(\varsigma,t)$ of infinitesimal thickness, 
where $\varsigma $ is arc length.
The velocity of the curve at the point $\bf{s}$ is 
${\bf{v}}_L = {\bf{v}}_{si} - \alpha {\bf{s}'}
\times {\bf{v}}_{si}$, where ${\bf{v}}_{si}$ is the self--induced velocity
(determined by a Biot--Savart integral over the entire vortex
configuration), ${\bf{s}'} \equiv d{\bf{s}'}/d\varsigma$ is the unit 
tangent vector at $\bf{s}$, and $\alpha$ is a dimensionless temperature
dependent friction parameter.  In the full expression ${\bf{v}}_L$ there is a second friction parameter, $\alpha'$, which we have neglected here since it is much smaller than $\alpha$~\citep{jackson_proukakis_09,berloff_youd_07,berloff_brachet_14}.  In superfluid helium, outside the
phase transition region (less than 1 percent from $T_c$), $\alpha$ is less
than unity and positive~\citep{jackson_proukakis_09,berloff_youd_07}. In atomic condensates, numerical simulations of vortex motion
based on the ZNG model have shown that \citep{jackson_proukakis_09}
$\alpha<0.03$ for $T/T_c<0.8$. The small value of $\alpha$ has been confirmed by an independent
calculation based on a classical field approach \citep{berloff_youd_07}.
The Biot--Savart model thus suggests that, instantaneously, 
the friction gives only a small contribution to the velocity of the
vortex. One expects the friction to be significant only on large enough
length scales and time scales, provided that its effects can accumulate.
For example, in the case of a single off-centered vortex in a 
harmonic trap precessing at finite temperature, the 
interaction of the vortex with the thermal cloud causes it to lose
energy and spiral out of the condensate, thus limiting its 
lifetime~\cite{allen_zaremba_13,fedichev_shlyapnikov_99,
schmidt_goral_03,duine_leurs_04,jackson_proukakis_09,
rooney_bradley_10}.  However, this decay may require many orbits in the trap
\citep{madarassy_barenghi_08,schmidt_goral_03,jackson_proukakis_09,
allen_zaremba_13}.

In conclusion, we have found that, on the typical short
length scales and time scales relevant to a vortex reconnection
in an atomic Bose--Einstein
condensate, the reconnection is essentially temperature independent,
despite the significant inhomogeneity of the thermal cloud in the vortex 
cores and near the boundary of the condensate. Since vortex reconnections
are essential ingredients of turbulence, our result suggests
that at least this rapid part of the dynamics is rather universal, and does
not depend on $T$, although the large scale motion of vortices does
depend on $T$.

AJA, NPP and CFB gratefully acknowledge funding from the EPSRC (Grant number: EP/I019413/1).


\begin{thebibliography}{36}
\expandafter\ifx\csname natexlab\endcsname\relax\def\natexlab#1{#1}\fi
\expandafter\ifx\csname bibnamefont\endcsname\relax
  \def\bibnamefont#1{#1}\fi
\expandafter\ifx\csname bibfnamefont\endcsname\relax
  \def\bibfnamefont#1{#1}\fi
\expandafter\ifx\csname citenamefont\endcsname\relax
  \def\citenamefont#1{#1}\fi
\expandafter\ifx\csname url\endcsname\relax
  \def\url#1{\texttt{#1}}\fi
\expandafter\ifx\csname urlprefix\endcsname\relax\def\urlprefix{URL }\fi
\providecommand{\bibinfo}[2]{#2}
\providecommand{\eprint}[2][]{\url{#2}}

\bibitem[{\citenamefont{Paoletti et~al.}(2010)\citenamefont{Paoletti, Fisher,
  and Lathrop}}]{Paoletti2010}
\bibinfo{author}{\bibfnamefont{M.~S.} \bibnamefont{Paoletti}},
  \bibinfo{author}{\bibfnamefont{M.~E.} \bibnamefont{Fisher}},
  \bibnamefont{and} \bibinfo{author}{\bibfnamefont{D.~P.}
  \bibnamefont{Lathrop}}, \bibinfo{journal}{Physica D}
  \textbf{\bibinfo{volume}{239}}, \bibinfo{pages}{1367} (\bibinfo{year}{2010}).

\bibitem[{\citenamefont{Barenghi et~al.}(2014)\citenamefont{Barenghi, Skrbek,
  and Sreenivasan}}]{Barenghi-Skrbek-Sreeni2014}
\bibinfo{author}{\bibfnamefont{C.~F.} \bibnamefont{Barenghi}},
  \bibinfo{author}{\bibfnamefont{L.}~\bibnamefont{Skrbek}}, \bibnamefont{and}
  \bibinfo{author}{\bibfnamefont{K.~R.} \bibnamefont{Sreenivasan}},
  \bibinfo{journal}{Proc. Nat. Acad. Sci. USA (Supp. 1)}
  \textbf{\bibinfo{volume}{111}}, \bibinfo{pages}{4647} (\bibinfo{year}{2014}).

\bibitem[{\citenamefont{Leadbeater et~al.}(2001)\citenamefont{Leadbeater,
  Winiecki, Samuels, Barenghi, and Adams}}]{Leadbeater2001}
\bibinfo{author}{\bibfnamefont{M.}~\bibnamefont{Leadbeater}},
  \bibinfo{author}{\bibfnamefont{T.}~\bibnamefont{Winiecki}},
  \bibinfo{author}{\bibfnamefont{D.~C.} \bibnamefont{Samuels}},
  \bibinfo{author}{\bibfnamefont{C.~F.}~\bibnamefont{Barenghi}}, \bibnamefont{and}
  \bibinfo{author}{\bibfnamefont{C.~S.} \bibnamefont{Adams}},
  \bibinfo{journal}{Phys. Rev. Lett.} \textbf{\bibinfo{volume}{86}},
  \bibinfo{pages}{1410} (\bibinfo{year}{2001}).

\bibitem[{\citenamefont{L'vov et~al.}(2008)\citenamefont{L'vov, Nazarenko, and
  Rudenko}}]{Lvov-Nazarenko-Rudenko2008}
\bibinfo{author}{\bibfnamefont{V.~S.} \bibnamefont{L'vov}},
  \bibinfo{author}{\bibfnamefont{S.~V.} \bibnamefont{Nazarenko}},
  \bibnamefont{and} \bibinfo{author}{\bibfnamefont{O.}~\bibnamefont{Rudenko}},
  \bibinfo{journal}{J. Low Temp. Phys.} \textbf{\bibinfo{volume}{153}},
  \bibinfo{pages}{140} (\bibinfo{year}{2008}).

\bibitem[{\citenamefont{Kursa et~al.}(2011)\citenamefont{Kursa, Bajer, and
  Lipniacki}}]{Kursa}
\bibinfo{author}{\bibfnamefont{M.}~\bibnamefont{Kursa}},
  \bibinfo{author}{\bibfnamefont{K.}~\bibnamefont{Bajer}}, \bibnamefont{and}
  \bibinfo{author}{\bibfnamefont{T.}~\bibnamefont{Lipniacki}},
  \bibinfo{journal}{Phys. Rev. B} \textbf{\bibinfo{volume}{83}},
  \bibinfo{pages}{014515} (\bibinfo{year}{2011}).

\bibitem[{\citenamefont{Kerr}(2011)}]{Kerr}
\bibinfo{author}{\bibfnamefont{R.~M.} \bibnamefont{Kerr}},
  \bibinfo{journal}{Phys. Rev. Lett.} \textbf{\bibinfo{volume}{106}},
  \bibinfo{pages}{224501} (\bibinfo{year}{2011}).

\bibitem[{\citenamefont{Leanhardt et~al.}(2002)\citenamefont{Leanhardt,
  G\"orlitz, Chikkatur, Kielpinski, Shin, Pritchard, and
  Ketterle}}]{leanhardt_gorlitz_02}
\bibinfo{author}{\bibfnamefont{A.~E.} \bibnamefont{Leanhardt}},
  \bibinfo{author}{\bibfnamefont{A.}~\bibnamefont{G\"orlitz}},
  \bibinfo{author}{\bibfnamefont{A.~P.} \bibnamefont{Chikkatur}},
  \bibinfo{author}{\bibfnamefont{D.}~\bibnamefont{Kielpinski}},
  \bibinfo{author}{\bibfnamefont{Y.}~\bibnamefont{Shin}},
  \bibinfo{author}{\bibfnamefont{D.~E.} \bibnamefont{Pritchard}},
  \bibnamefont{and} \bibinfo{author}{\bibfnamefont{W.}~\bibnamefont{Ketterle}},
  \bibinfo{journal}{Phys. Rev. Lett.} \textbf{\bibinfo{volume}{89}},
  \bibinfo{pages}{190403} (\bibinfo{year}{2002}).

\bibitem[{\citenamefont{Raman et~al.}(2001)\citenamefont{Raman, Abo-Shaeer,
  Vogels, Xu, and Ketterle}}]{raman_aboshaeer_01}
\bibinfo{author}{\bibfnamefont{C.}~\bibnamefont{Raman}},
  \bibinfo{author}{\bibfnamefont{J.~R.} \bibnamefont{Abo-Shaeer}},
  \bibinfo{author}{\bibfnamefont{J.~M.} \bibnamefont{Vogels}},
  \bibinfo{author}{\bibfnamefont{K.}~\bibnamefont{Xu}}, \bibnamefont{and}
  \bibinfo{author}{\bibfnamefont{W.}~\bibnamefont{Ketterle}},
  \bibinfo{journal}{Phys. Rev. Lett.} \textbf{\bibinfo{volume}{87}},
  \bibinfo{pages}{210402} (\bibinfo{year}{2001}).

\bibitem[{\citenamefont{Anderson et~al.}(2001)\citenamefont{Anderson, Haljan,
  Regal, Feder, Collins, Clark, and Cornell}}]{anderson_haljan_01}
\bibinfo{author}{\bibfnamefont{B.~P.} \bibnamefont{Anderson}},
  \bibinfo{author}{\bibfnamefont{P.~C.} \bibnamefont{Haljan}},
  \bibinfo{author}{\bibfnamefont{C.~A.} \bibnamefont{Regal}},
  \bibinfo{author}{\bibfnamefont{D.~L.} \bibnamefont{Feder}},
  \bibinfo{author}{\bibfnamefont{L.~A.} \bibnamefont{Collins}},
  \bibinfo{author}{\bibfnamefont{C.~W.} \bibnamefont{Clark}}, \bibnamefont{and}
  \bibinfo{author}{\bibfnamefont{E.~A.} \bibnamefont{Cornell}},
  \bibinfo{journal}{Phys. Rev. Lett.} \textbf{\bibinfo{volume}{86}},
  \bibinfo{pages}{2926} (\bibinfo{year}{2001}).

\bibitem[{\citenamefont{Weiler et~al.}(2008)\citenamefont{Weiler, Neely,
  Scherer, Bradley, Davis, and Anderson}}]{weiler_neely_08}
\bibinfo{author}{\bibfnamefont{C.}~\bibnamefont{Weiler}},
  \bibinfo{author}{\bibfnamefont{T.~W.} \bibnamefont{Neely}},
  \bibinfo{author}{\bibfnamefont{D.~R.} \bibnamefont{Scherer}},
  \bibinfo{author}{\bibfnamefont{A.~S.} \bibnamefont{Bradley}},
  \bibinfo{author}{\bibfnamefont{M.~J.} \bibnamefont{Davis}}, \bibnamefont{and}
  \bibinfo{author}{\bibfnamefont{B.~P.} \bibnamefont{Anderson}},
  \bibinfo{journal}{Nature} \textbf{\bibinfo{volume}{455}},
  \bibinfo{pages}{948} (\bibinfo{year}{2008}).

\bibitem[{\citenamefont{Freilich et~al.}(2010)\citenamefont{Freilich, Bianchi,
  Kaufman, Langin, and Hall}}]{freilich_bianchi_10}
\bibinfo{author}{\bibfnamefont{D.~V.} \bibnamefont{Freilich}},
  \bibinfo{author}{\bibfnamefont{D.~M.} \bibnamefont{Bianchi}},
  \bibinfo{author}{\bibfnamefont{A.~M.} \bibnamefont{Kaufman}},
  \bibinfo{author}{\bibfnamefont{T.~K.} \bibnamefont{Langin}},
  \bibnamefont{and} \bibinfo{author}{\bibfnamefont{D.~S.} \bibnamefont{Hall}},
  \bibinfo{journal}{Science} \textbf{\bibinfo{volume}{329}},
  \bibinfo{pages}{1182} (\bibinfo{year}{2010}).

\bibitem[{\citenamefont{Davis et~al.}(2009)\citenamefont{Davis,
  Carretero-Gonz\'alez, Shi, Law, Kevrekidis, and
  Anderson}}]{davis_carretero_09}
\bibinfo{author}{\bibfnamefont{M.~C.} \bibnamefont{Davis}},
  \bibinfo{author}{\bibfnamefont{R.}~\bibnamefont{Carretero-Gonz\'alez}},
  \bibinfo{author}{\bibfnamefont{Z.}~\bibnamefont{Shi}},
  \bibinfo{author}{\bibfnamefont{K.~J.~H.} \bibnamefont{Law}},
  \bibinfo{author}{\bibfnamefont{P.~G.} \bibnamefont{Kevrekidis}},
  \bibnamefont{and} \bibinfo{author}{\bibfnamefont{B.~P.}
  \bibnamefont{Anderson}}, \bibinfo{journal}{Phys. Rev. A}
  \textbf{\bibinfo{volume}{80}}, \bibinfo{pages}{023604}
  (\bibinfo{year}{2009}).

\bibitem[{\citenamefont{Madison et~al.}(2000)\citenamefont{Madison, Chevy,
  Wohlleben, and Dalibard}}]{madison_chevy_00}
\bibinfo{author}{\bibfnamefont{K.~W.} \bibnamefont{Madison}},
  \bibinfo{author}{\bibfnamefont{F.}~\bibnamefont{Chevy}},
  \bibinfo{author}{\bibfnamefont{W.}~\bibnamefont{Wohlleben}},
  \bibnamefont{and} \bibinfo{author}{\bibfnamefont{J.}~\bibnamefont{Dalibard}},
  \bibinfo{journal}{Phys. Rev. Lett.} \textbf{\bibinfo{volume}{84}},
  \bibinfo{pages}{806} (\bibinfo{year}{2000}).

\bibitem{baggaley_14} A. W. Baggaley (2014),
\bibinfo{note}{arXiv:1403.8121 [physics.flu-dyn]}, \eprint{1403.8121}. 

\bibitem[{\citenamefont{Neely et~al.}(2013)\citenamefont{Neely, Bradley,
  Samson, Rooney, Wright, Law, Carretero-Gonz\'alez, Kevrekidis, Davis, and
  Anderson}}]{neely_bradley_13}
\bibinfo{author}{\bibfnamefont{T.~W.} \bibnamefont{Neely}},
  \bibinfo{author}{\bibfnamefont{A.~S.} \bibnamefont{Bradley}},
  \bibinfo{author}{\bibfnamefont{E.~C.} \bibnamefont{Samson}},
  \bibinfo{author}{\bibfnamefont{S.~J.} \bibnamefont{Rooney}},
  \bibinfo{author}{\bibfnamefont{E.~M.} \bibnamefont{Wright}},
  \bibinfo{author}{\bibfnamefont{K.~J.~H.} \bibnamefont{Law}},
  \bibinfo{author}{\bibfnamefont{R.}~\bibnamefont{Carretero-Gonz\'alez}},
  \bibinfo{author}{\bibfnamefont{P.~G.} \bibnamefont{Kevrekidis}},
  \bibinfo{author}{\bibfnamefont{M.~J.} \bibnamefont{Davis}}, \bibnamefont{and}
  \bibinfo{author}{\bibfnamefont{B.~P.} \bibnamefont{Anderson}},
  \bibinfo{journal}{Phys. Rev. Lett.} \textbf{\bibinfo{volume}{111}},
  \bibinfo{pages}{235301} (\bibinfo{year}{2013}).

\bibitem{wilson_samson_13}
K. E. Wilson, E. C. Samson, Z. L. Newman, T. W. Neely, and B. P. Anderson, 
{\it{Annual Review of Cold Atoms and Molecules}} {\bf{1}}, pp. 261-298. (2013).



\bibitem[{\citenamefont{Kwon et~al.}(2014)\citenamefont{Kwon, Moon, Choi, Seo,
  and Shin}}]{kwon_moon_14}
\bibinfo{author}{\bibfnamefont{W.~J.} \bibnamefont{Kwon}},
  \bibinfo{author}{\bibfnamefont{G.}~\bibnamefont{Moon}},
  \bibinfo{author}{\bibfnamefont{J.}~\bibnamefont{Choi}},
  \bibinfo{author}{\bibfnamefont{S.~W.} \bibnamefont{Seo}}, \bibnamefont{and}
  \bibinfo{author}{\bibfnamefont{Y.}~\bibnamefont{Shin}}
  (\bibinfo{year}{2014}), \bibinfo{note}{arXiv:1403.4658 [cond-mat.quant-gas]},
  \eprint{1403.4658}.

\bibitem[{\citenamefont{Henn et~al.}(2009)\citenamefont{Henn, Seman, Roati,
  Magalh\~aes, and Bagnato}}]{henn_seman_09}
\bibinfo{author}{\bibfnamefont{E.~A.~L.} \bibnamefont{Henn}},
  \bibinfo{author}{\bibfnamefont{J.~A.} \bibnamefont{Seman}},
  \bibinfo{author}{\bibfnamefont{G.}~\bibnamefont{Roati}},
  \bibinfo{author}{\bibfnamefont{K.~M.~F.} \bibnamefont{Magalh\~aes}},
  \bibnamefont{and} \bibinfo{author}{\bibfnamefont{V.~S.}
  \bibnamefont{Bagnato}}, \bibinfo{journal}{Phys. Rev. Lett.}
  \textbf{\bibinfo{volume}{103}}, \bibinfo{pages}{045301}
  (\bibinfo{year}{2009}).

\bibitem{white_anderson_14}
A. C. White, B. P. Anderson, and V. S. Bagnato, Proc. Nat. Acad. Sci USA (Suppl. 1) {\bf 111}, 4719 (2014).
 

%
%
%


\bibitem[{\citenamefont{Fedichev and
  Shlyapnikov}(1999)}]{fedichev_shlyapnikov_99}
\bibinfo{author}{\bibfnamefont{P.~O.} \bibnamefont{Fedichev}} \bibnamefont{and}
  \bibinfo{author}{\bibfnamefont{G.~V.} \bibnamefont{Shlyapnikov}},
  \bibinfo{journal}{Phys. Rev. A} \textbf{\bibinfo{volume}{60}},
  \bibinfo{pages}{R1779} (\bibinfo{year}{1999}).

\bibitem[{\citenamefont{Schmidt et~al.}(2003)\citenamefont{Schmidt, Goral,
  Gajda, and Rzazewski}}]{schmidt_goral_03}
\bibinfo{author}{\bibfnamefont{H.}~\bibnamefont{Schmidt}},
  \bibinfo{author}{\bibfnamefont{F.}~\bibnamefont{Goral},
  \bibfnamefont{K.~Floegel}},
  \bibinfo{author}{\bibfnamefont{M.}~\bibnamefont{Gajda}}, \bibnamefont{and}
  \bibinfo{author}{\bibfnamefont{K.}~\bibnamefont{Rzazewski}},
  \bibinfo{journal}{J. Opt. B: Quantum Semiclass} \textbf{\bibinfo{volume}{5}},
  \bibinfo{pages}{S96} (\bibinfo{year}{2003}).

\bibitem[{\citenamefont{Duine et~al.}(2004)\citenamefont{Duine, Leurs, and
  Stoof}}]{duine_leurs_04}
\bibinfo{author}{\bibfnamefont{R.~A.} \bibnamefont{Duine}},
  \bibinfo{author}{\bibfnamefont{B.~W.~A.} \bibnamefont{Leurs}},
  \bibnamefont{and} \bibinfo{author}{\bibfnamefont{H.~T.~C.}
  \bibnamefont{Stoof}}, \bibinfo{journal}{Phys. Rev. A}
  \textbf{\bibinfo{volume}{69}}, \bibinfo{pages}{053623}
  (\bibinfo{year}{2004}).

\bibitem[{\citenamefont{Jackson et~al.}(2009)\citenamefont{Jackson, Proukakis,
  Barenghi, and Zaremba}}]{jackson_proukakis_09}
\bibinfo{author}{\bibfnamefont{B.}~\bibnamefont{Jackson}},
  \bibinfo{author}{\bibfnamefont{N.~P.} \bibnamefont{Proukakis}},
  \bibinfo{author}{\bibfnamefont{C.~F.} \bibnamefont{Barenghi}},
  \bibnamefont{and} \bibinfo{author}{\bibfnamefont{E.}~\bibnamefont{Zaremba}},
  \bibinfo{journal}{Phys. Rev. A} \textbf{\bibinfo{volume}{79}},
  \bibinfo{pages}{053615} (\bibinfo{year}{2009}).

\bibitem[{\citenamefont{Rooney et~al.}(2010)\citenamefont{Rooney, Bradley, and
  Blakie}}]{rooney_bradley_10}
\bibinfo{author}{\bibfnamefont{S.~J.} \bibnamefont{Rooney}},
  \bibinfo{author}{\bibfnamefont{A.~S.} \bibnamefont{Bradley}},
  \bibnamefont{and} \bibinfo{author}{\bibfnamefont{P.~B.}
  \bibnamefont{Blakie}}, \bibinfo{journal}{Phys. Rev. A}
  \textbf{\bibinfo{volume}{81}}, \bibinfo{pages}{023630}
  (\bibinfo{year}{2010}).

\bibitem[{\citenamefont{Allen et~al.}(2013)\citenamefont{Allen, Zaremba,
  Barenghi, and Proukakis}}]{allen_zaremba_13}
\bibinfo{author}{\bibfnamefont{A.~J.} \bibnamefont{Allen}},
  \bibinfo{author}{\bibfnamefont{E.}~\bibnamefont{Zaremba}},
  \bibinfo{author}{\bibfnamefont{C.~F.} \bibnamefont{Barenghi}},
  \bibnamefont{and} \bibinfo{author}{\bibfnamefont{N.~P.}
  \bibnamefont{Proukakis}}, \bibinfo{journal}{Phys. Rev. A}
  \textbf{\bibinfo{volume}{87}}, \bibinfo{pages}{013630}
  (\bibinfo{year}{2013}).

\bibitem[{\citenamefont{Zaremba et~al.}(1999)\citenamefont{Zaremba, Nikuni, and
  Griffin}}]{zaremba_nikuni_99}
\bibinfo{author}{\bibfnamefont{E.}~\bibnamefont{Zaremba}},
  \bibinfo{author}{\bibfnamefont{T.}~\bibnamefont{Nikuni}}, \bibnamefont{and}
  \bibinfo{author}{\bibfnamefont{A.}~\bibnamefont{Griffin}},
  \bibinfo{journal}{Journal of Low Temperature Physics}
  \textbf{\bibinfo{volume}{116}}, \bibinfo{pages}{277} (\bibinfo{year}{1999}).
  

\bibitem[{\citenamefont{Griffin et~al.}(2009)\citenamefont{Griffin, Nikuni, and
  Zaremba}}]{griffin_nikuni_book_09}
\bibinfo{author}{\bibfnamefont{A.}~\bibnamefont{Griffin}},
  \bibinfo{author}{\bibfnamefont{T.}~\bibnamefont{Nikuni}}, \bibnamefont{and}
  \bibinfo{author}{\bibfnamefont{E.}~\bibnamefont{Zaremba}},
  {\it{{Bose-condensed gases at finite temperatures}}}
  (\bibinfo{publisher}{Cambridge University Press}, \bibinfo{year}{2009}).

\bibitem[{\citenamefont{Choi et~al.}(1998)\citenamefont{Choi, Morgan, and
  Burnett}}]{choi_morgan_98}
\bibinfo{author}{\bibfnamefont{S.}~\bibnamefont{Choi}},
  \bibinfo{author}{\bibfnamefont{S.~A.} \bibnamefont{Morgan}},
  \bibnamefont{and} \bibinfo{author}{\bibfnamefont{K.}~\bibnamefont{Burnett}},
 \bibinfo{journal}{Phys. Rev. A} \textbf{\bibinfo{volume}{57}},
  \bibinfo{pages}{4057} (\bibinfo{year}{1998}).

\bibitem[{\citenamefont{Pitaevskii and
  Stringari}(2003)}]{pitaevskii_stringari_book_03}
\bibinfo{author}{\bibfnamefont{L.~P.} \bibnamefont{Pitaevskii}}
  \bibnamefont{and}
  \bibinfo{author}{\bibfnamefont{S.}~\bibnamefont{Stringari}},
  {\it{\bibinfo{title}{Bose-Einstein Condensation}}}
  (\bibinfo{publisher}{Oxford University Press}, \bibinfo{address}{Great
  Clarendon Street, Oxford}, \bibinfo{year}{2003}).

\bibitem[{\citenamefont{Pethick and Smith}(2002)}]{pethick_smith_book_02}
\bibinfo{author}{\bibfnamefont{C.}~\bibnamefont{Pethick}} \bibnamefont{and}
  \bibinfo{author}{\bibfnamefont{H.}~\bibnamefont{Smith}},
  {\it{\bibinfo{title}{Bose-Einstein condensation in dilute gases}}}
  (\bibinfo{publisher}{Cambridge University Press}, \bibinfo{year}{2002}).

\bibitem{crow} 
S. C. Crow, AIAA J. {\bf{8}}, 2172 (1970).


\bibitem[{\citenamefont{White et~al.}(2010)\citenamefont{White, Barenghi,
  Proukakis, Youd, and Wacks}}]{white2010}
\bibinfo{author}{\bibfnamefont{A.~C.} \bibnamefont{White}},
  \bibinfo{author}{\bibfnamefont{C.~F.} \bibnamefont{Barenghi}},
  \bibinfo{author}{\bibfnamefont{N.~P.}~\bibnamefont{Proukakis}},
  \bibinfo{author}{\bibfnamefont{A.~J.} \bibnamefont{Youd}}, \bibnamefont{and}
  \bibinfo{author}{\bibfnamefont{D.~H.} \bibnamefont{Wacks}},
  \bibinfo{journal}{Phys. Rev. Lett.} \textbf{\bibinfo{volume}{104}},
  \bibinfo{pages}{075301} (\bibinfo{year}{2010}).


\bibitem{jackson_proukakis_08}  N.~P. Proukakis and B.~J. Jackson, J. Phys. B: At. Mol. Opt. Phys. {\bf{41}}, 203002 (2008).


\bibitem[{\citenamefont{Tsubota et~al.}(2002)\citenamefont{Tsubota, Kasamatsu,
  and Ueda}}]{tsubota_kasamatsu_02}
\bibinfo{author}{\bibfnamefont{M.}~\bibnamefont{Tsubota}},
  \bibinfo{author}{\bibfnamefont{K.}~\bibnamefont{Kasamatsu}},
  \bibnamefont{and} \bibinfo{author}{\bibfnamefont{M.}~\bibnamefont{Ueda}},
  \bibinfo{journal}{Phys. Rev. A} \textbf{\bibinfo{volume}{65}},
  \bibinfo{pages}{023603} (\bibinfo{year}{2002}).

\bibitem{proukakis_parker_04} N. P. Proukakis, N. G. Parker, C. F. Barenghi, and and C. S. Adams, Phys. Rev. Lett. {\bf{93}}, 130408 (2004).


\bibitem[{\citenamefont{Madarassy and Barenghi}(2008)}]{madarassy_barenghi_08}
\bibinfo{author}{\bibfnamefont{E.}~\bibnamefont{Madarassy}} \bibnamefont{and}
  \bibinfo{author}{\bibfnamefont{C.}~\bibnamefont{Barenghi}},
  \bibinfo{journal}{Journal of Low Temperature Physics}
  \textbf{\bibinfo{volume}{152}}, \bibinfo{pages}{122} (\bibinfo{year}{2008}).

 \bibitem[{\citenamefont{Zuccher et~al.}(2012)\citenamefont{Zuccher, Caliari,
  Baggaley, and Barenghi}}]{zuccher_caliari_12}
\bibinfo{author}{\bibfnamefont{S.}~\bibnamefont{Zuccher}},
  \bibinfo{author}{\bibfnamefont{M.}~\bibnamefont{Caliari}},
  \bibinfo{author}{\bibfnamefont{A.~W.} \bibnamefont{Baggaley}},
  \bibnamefont{and} \bibinfo{author}{\bibfnamefont{C.~F.}
  \bibnamefont{Barenghi}}, \bibinfo{journal}{Physics of Fluids }
  \textbf{\bibinfo{volume}{24}}, \bibinfo{eid}{125108} (\bibinfo{year}{2012}).
 
\bibitem{bao_etal_2006}
 W. Bao, Q. Du, and Y. Zhang, 
 {SIAM} J. Appl. Math. {\bf{66}}, No. 3, 758--786 (2006). 


\bibitem{penckwitt_ballagh_02} A.~A. Penckwitt, R.~J. Ballagh, and C.~W. Gardiner, Phys. Rev. Lett. {\bf{89}}, 260402 (2002).

\bibitem{kasamatsu_tsubota_03} K. Kasamatsu, M. Tsubota, and M. Ueda, Phys. Rev. A {\bf{67}}, 033610 (2003).


\bibitem{allen_parker_14}
 A. J. Allen, N. G. Parker, N. P. Proukakis, and C. F. Barenghi, 
 Phys. Rev. A {\bf{89}}, 025602 (2014). 



\bibitem[{\citenamefont{Hussain and Duraisamy}(2011)}]{hussain_duraisamy_11}
\bibinfo{author}{\bibfnamefont{F.}~\bibnamefont{Hussain}} \bibnamefont{and}
  \bibinfo{author}{\bibfnamefont{K.}~\bibnamefont{Duraisamy}},
  \bibinfo{journal}{Physics of Fluids (1994-present)}
  \textbf{\bibinfo{volume}{23}}, \bibinfo{eid}{021701} (\bibinfo{year}{2011}).


\bibitem{Hanninen}
R. H\"{a}nninen and A.W. Baggaley,
Proc. Nat. Acad. Sci USA (Suppl. 1) {\bf 111}, 4667 (2014).


\bibitem{berloff_youd_07}
N. G. Berloff and A. J. Youd, 
Phys. Rev. Lett. {\bf{99}}, 145301 (2007).

\bibitem{berloff_brachet_14}
N. G. Berloff, M. Brachet, and N. P. Proukakis, Proc. Nat. Acad. Sci USA (Suppl. 1) {\bf{111}}, 4657 (2014).

\end{thebibliography}

\end{document}